\documentclass[11pt]{article}
\usepackage[T1]{fontenc}
\usepackage[utf8]{inputenc}
\usepackage{lmodern}
\usepackage[a4paper,margin=1in]{geometry}
\usepackage{microtype}
\usepackage{amsmath,amssymb,amsthm,bm}
\usepackage{graphicx,booktabs,array}
\usepackage{float}
\floatstyle{ruled}
\newfloat{algorithm}{tbp}{loa}
\floatname{algorithm}{Algorithm}
\usepackage{placeins}
\usepackage[hidelinks]{hyperref}
\hypersetup{pdftitle={Relative-Entropy Geometry and Convergence in Nonconvex Wiretap Optimization},
pdfauthor={Farzin Salek and Masahito Hayashi}}
\usepackage{url}
\newtheorem{theorem}{Theorem}[section]
\newtheorem{proposition}[theorem]{Proposition}

\theoremstyle{definition}

\newtheorem{example}[theorem]{Example}
\theoremstyle{remark}

\DeclareMathOperator{\Tr}{Tr}
\DeclareMathOperator{\supp}{supp}

\DeclareMathOperator*{\argmin}{arg\,min}
\newcommand{\cX}{\mathcal X}

\newcommand{\cU}{\mathcal U}
\newcommand{\cH}{\mathcal H}
\newcommand{\cD}{\mathcal D}

\begin{document}
\title{Relative-Entropy Geometry and Convergence in \\ Nonconvex Wiretap Optimization}
\author{\begin{minipage}{.95\textwidth}\centering
Farzin Salek$^{1,2,3,4}$\thanks{\texttt{farzin.salek@gmail.com}}
\quad and\quad
Masahito Hayashi$^{5,6,7}$\thanks{\texttt{hmasahito@cuhk.edu.cn}}\\[1ex]
\small
$^1$Department of Combinatorics and Optimization, Faculty of Mathematics,\\
University of Waterloo, Waterloo, ON N2L 3G1, Canada\\[.3ex]
$^2$Perimeter Institute for Theoretical Physics, Waterloo, ON N2L 2Y5, Canada\\[.3ex]
$^3$Institute for Quantum Computing, University of Waterloo, Waterloo, ON N2L 3G1, Canada\\[.3ex]
$^4$Dahlem Center for Complex Quantum Systems, Freie Universit\"at Berlin, Berlin, Germany\\[.3ex]
$^5$School of Data Science, The Chinese University of Hong Kong, Shenzhen,\\
Longgang District, Shenzhen 518172, China\\[.3ex]
$^6$International Quantum Academy, Futian District, Shenzhen 518048, China\\[.3ex]
$^7$Graduate School of Mathematics, Nagoya University, Nagoya 464-8602, Japan
\end{minipage}}
\date{}
\maketitle
\begin{abstract}
We develop and analyze an entropic descent method for a nonconvex optimization problem on probability simplices arising in classical and quantum secure communication. The optimization variables describe a randomized encoder, allowing its input distribution and stochastic preprocessing to be optimized jointly. 
An exact relative-entropy identity yields a channel-independent majorization bound and an explicit mirror-descent iteration with a fixed unit step size. 
We establish quantitative descent, convergence of objective
values, and stationarity of accumulation points on their support
faces for arbitrary finite classical channels,
without assuming degradedness or less noisiness.
For strictly positive transition matrices, a square-root parametrization and a Kurdyka--\L{}ojasiewicz argument establish finite length of the transformed trajectory and convergence of the entire sequence, including limits on the simplex boundary. For less-noisy channels, a restricted convex formulation admits global convergence with an explicit \(O(1/t)\) objective-error bound. We extend the relative-entropy construction to noncommuting quantum outputs and derive monotone alternating updates for covariant encoders, including correlated preprocessing over a fixed block. Numerical experiments compare the method with analytic benchmarks and constrained optimization, reveal multiple attracting values, and demonstrate strict improvements from stochastic preprocessing.
\end{abstract}

\section{Introduction}
\label{sec:introduction}
\subsection{Formulation of secrecy capacity}
The secrecy capacity of a discrete memoryless wiretap channel is a
basic example of an information-theoretic quantity whose operational
characterization is simpler than its numerical evaluation. For a main
channel $W_Y:\mathcal X\to\mathcal P(\mathcal Y)$ and an eavesdropper
channel $W_Z:\mathcal X\to\mathcal P(\mathcal Z)$, the
Csisz\'ar--K\"orner formula is
\begin{equation}
 C_{\mathrm s}=\max_{p_{UX}}\{I(U;Y)-I(U;Z)\},
 \qquad U\longrightarrow X\longrightarrow(Y,Z).
 \label{eq:secrecy-capacity}
\end{equation}
Here $U$ labels the information-bearing signals, and the conditional
distribution $p_{X|U}$ is the stochastic encoder, also called channel
prefixing~\cite{WynerWiretap,CsiszarKorner}. One may take
$|\mathcal U|\le |\mathcal X|$; fixing this alphabet turns
\eqref{eq:secrecy-capacity} into an optimization over a finite simplex.
Optimizing only the input distribution with $U=X$ can give a strictly
smaller rate. Optimizing the full joint law, however, is generally
nonconvex.

Auxiliary distributions and encoder randomness also enter quantum
broadcast coding with common, individualized, and confidential
messages~\cite{SalekBroadcast2020}, three-receiver quantum broadcast
coding~\cite{SalekBroadcast2026}, and broadcast channels with mutually
confidential messages~\cite{BelzigConfidential2026}. These operational
results motivate numerical methods that optimize stochastic encoders
explicitly. The present paper concerns the wiretap objective
\eqref{eq:secrecy-capacity} and its classical--quantum analogue; we do
not optimize the larger broadcast rate regions.

The difficulty of optimizing auxiliary distributions also arises in
classical broadcast theory: Gohari and Anantharam~\cite{GohariAnantharam2012}
developed perturbation methods to establish cardinality bounds and
computability of Marton's inner bound.

The distinction between optimizing an input law and optimizing a joint
encoder law matters for the geometry. With a fixed channel,
$p_X\mapsto I(X;Y)$ is concave. In contrast, when $p_{UX}$ varies,
the effective channel from $U$ to $Y$ also varies, and $I(U;Y)$ is
not generally concave in $p_{UX}$. Even the restricted difference
$I(X;Y)-I(X;Z)$ need not be concave. Multiple attracting stationary
values can therefore occur, and monotonic ascent alone does not imply
that an iteration computes the global capacity.

For finite-dimensional quantum-input channels, Wilde, Tomamichel,
and Berta~\cite{WildeTomamichelBerta2017} derived converse bounds for
private communication using privacy tests, while Leung and
Smith~\cite{LeungSmithContinuity2009} established continuity bounds for
channel capacities, including private capacity, under channel
perturbations. Li, Winter, Zou, and Guo~\cite{LiWinterZouGuo2009}
demonstrated nonadditivity of private capacity, highlighting the role
of joint coding across channel uses.

\subsection{Related algorithms and contribution}
Blahut--Arimoto-type algorithms exploit entropy identities rather than
treating the capacity formula as an arbitrary nonlinear objective.
Yasui, Suko, and Matsushima developed a globally convergent algorithm
for less-noisy wiretap channels~\cite{Yasui}; computation for
more-capable channel pairs was studied by Gowtham and
Thangaraj~\cite{GowthamThangaraj2008}. Quantum capacity
algorithms based on relative entropy were developed by Ramakrishnan,
Iten, Scholz, and Berta~\cite{RISB}. The broader iterative frameworks
of Hayashi~\cite{Haya} and Hayashi and Liu~\cite{HaLi}, and the
Bregman proximal interpretation of He, Saunderson, and
Fawzi~\cite{HeFawziBA}, provide the relevant algorithmic setting.
The preprint version of Hayashi~\cite{Haya} also gave an update for
general wiretap channels, with convergence conditions verified for a
separate degraded-channel formulation.

In this paper, we introduce a different update rule that jointly
optimizes the auxiliary distribution and the stochastic encoder. We prove
that the objective does not increase at any iteration, with a quantitative
descent bound and strict decrease unless the iterate is fixed. This
guarantee holds for arbitrary finite classical wiretap channel pairs,
without any channel-ordering assumption. 
We prove
convergence of its objective values and stationarity of every
accumulation point relative to its support face. A counterexample
explains why this does not imply convergence to a local optimum.
For strictly positive classical transition probabilities, a
Kurdyka--\L{}ojasiewicz argument proves convergence of the entire
sequence, including limits on the boundary.
When the main channel is less noisy, restriction to $U=X$ gives
convergence to a global optimizer with an explicit $O(1/t)$ value bound.
We extend the algorithm to wiretap channels with noncommuting quantum outputs and establish a quantitative descent guarantee in this setting. We also derive alternating updates for the covariant preprocessing distributions of Hayashi and Salek~\cite{SalekWiretap}.
Numerical experiments compare the
iteration with analytic formulas, binary convex-envelope benchmarks,
and direct constrained optimization, and explicitly measure gains
from channel prefixing.

\subsection{Channel ordering and the scope of capacity claims}
\label{subsec:ordering}
The classical main channel is \emph{more capable} than Eve's channel
if $I(X;Y)\ge I(X;Z)$ for every input law, and \emph{less noisy}
if $I(U;Y)\ge I(U;Z)$ for every joint law $p_{UX}$.
Stochastic degradedness implies less noisiness, which implies more
capability~\cite{KornerMarton,ElGamalKim}.
Writing
\begin{equation}
 f(p_X)=I(X;Y)-I(X;Z),
\end{equation}
the chain rule gives
\begin{equation}
 I(U;Y)-I(U;Z)
 =f(p_X)-\sum_u p_U(u)f(p_{X|U=u}).
 \label{eq:prefixing-gap}
\end{equation}
Thus more capability makes $U=X$ optimal, but does not in general
make $f$ concave. Less noisiness is equivalent to concavity of $f$:
\eqref{eq:prefixing-gap} is precisely the Jensen gap for any finite
mixture of input laws~\cite{vanDijk,OzelUlukusMoreCapable}.
In the less-noisy case,
\begin{equation}
 C_{\mathrm s}=\max_{p_X} f(p_X)
 \label{eq:Cs-more-capable}
\end{equation}
is consequently a concave maximization problem. The full
$p_{UX}$ parametrization need not become concave; the global guarantee
in this paper concerns the restricted $U=X$ iteration.

For a classical--quantum (cq) wiretap channel, the two marginal
channels assign density operators $\rho_x^B$ and $\sigma_x^E$ to
each classical input $x$. Its one-use private information is
\begin{equation}
 P^{(1)}(W)=\max_{p_{UX}}\{I(U;B)-I(U;E)\},
 \qquad |\mathcal U|\le |\mathcal X|.
 \label{eq:cq-private-info}
\end{equation}
The operational private capacity is in general
\begin{equation}
 C_{\mathrm p}(W)=\lim_{n\to\infty}\frac1n
 P^{(1)}(W^{\otimes n}),
 \label{eq:cq-regularization}
\end{equation}
and nonadditivity can occur even if only one receiver has quantum
outputs~\cite{CWY,Devetak,TikkuBertaRenes}.
The one-use cq more-capable and less-noisy conditions are defined by
replacing $Y,Z$ with $B,E$ in the preceding inequalities. They imply,
respectively, removal of $U$ from \eqref{eq:cq-private-info} and
concavity of $p_X\mapsto I(X;B)-I(X;E)$, by the same chain-rule
argument. These statements concern $P^{(1)}$.
Watanabe's stronger less-noisy condition involves every
blocklength~\cite{Watanabe}, whereas we use the one-use
condition~\cite{TikkuBertaRenes}. For degradable cq wiretap channels,
\eqref{eq:cq-regularization} reduces to one use~\cite{WinterCQConverse}.

\subsection{Dependence on the marginal channels}
For the unassisted, memoryless wiretap task considered here, both the
objective and the operational reliability and secrecy conditions
depend only on the two marginal channels. The coupling between the
receivers' outputs can therefore be chosen conveniently. For example,
one may use $W_Y(y|x)W_Z(z|x)$ in the classical model and
$\rho_x^B\otimes\sigma_x^E$ in the cq model. Physical degradedness
of a particular coupling is a distinct notion from stochastic
degradedness of the marginal channels.

All logarithms in the analysis and the update rules are natural.
Numerical rates are converted to bits by division by $\ln 2$.
The convention $0\ln0=0$ is used for entropies; logarithms of output
states are taken on their relevant supports.

\section{A variational functional and its multiplicative iteration}
\label{sec:classical-algorithm}

Write
$W_Y(y|x)=p_{Y|X}(y|x)$ and $W_Z(z|x)=p_{Z|X}(z|x)$.
For a joint law $p=p_{UX}$, let
\[
\begin{aligned}
p_U(u)&=\sum_xp(u,x),\qquad
p_{UA}(u,a)=\sum_xp(u,x)W_A(a|x),\\
p_A(a)&=\sum_up_{UA}(u,a),\qquad A\in\{Y,Z\}.
\end{aligned}
\]
When $p_U(u)>0$, write $p_{A|u}(a)=p_{UA}(u,a)/p_U(u)$.
We use $0\log0=0$ and omit output symbols unreachable from the
allowed inputs.

It is useful to allow a specified initial support
$\mathcal S\subseteq\mathcal U\times\mathcal X$ and to write
$\Delta_{\mathcal S}$ for the corresponding probability simplex.
The general algorithm uses $\mathcal S=\mathcal U\times\mathcal X$;
the restriction $U=X$ uses the diagonal support. We initialize
$p^{(0)}(u,x)>0$ on $\mathcal S$. Every finite iterate remains
positive there. Scores are evaluated only on this allowed support;
in particular, no conditional distribution is assigned to an unused
$U$-row merely to evaluate the algorithm.

\subsection{The objective and its score}

Define
\begin{equation}
F(p):=I(U;Z)_p-I(U;Y)_p.
\label{eq:F-MI-def}
\end{equation}
For an auxiliary alphabet large enough to represent the
Csisz\'ar--K\"orner optimization,
\begin{equation}
C=-\min_{p\in\Delta_{\mathcal U\times\mathcal X}}F(p).
\label{eq:C-as-min-F}
\end{equation}
For $p(u,x)>0$, define
\begin{align}
\Omega[p](u,x)
&:=\sum_zW_Z(z|x)\log\frac{p_{Z|u}(z)}{p_Z(z)}
 -\sum_yW_Y(y|x)\log\frac{p_{Y|u}(y)}{p_Y(y)}.
\label{eq:Omega-def}
\end{align}
Every term with a nonzero channel coefficient is finite at such a
coordinate. Summing over the joint law gives
\begin{equation}
F(p)=\langle p,\Omega[p]\rangle
    :=\sum_{(u,x)\in\supp p}p(u,x)\Omega[p](u,x).
\label{eq:F-as-Omega}
\end{equation}
The score is also the derivative of the objective on the simplex.
Indeed, cancellation of the $U$-entropy terms gives
\begin{align}
F(p)
={}&\sum_{u,z}p_{UZ}(u,z)\log p_{UZ}(u,z)
 -\sum_zp_Z(z)\log p_Z(z)\nonumber\\
&-\sum_{u,y}p_{UY}(u,y)\log p_{UY}(u,y)
 +\sum_yp_Y(y)\log p_Y(y).
\label{eq:F-entropy-form}
\end{align}
Differentiating this expression shows that, for a tangent direction
$h$ supported on $\mathcal S$ with $\sum_{u,x}h(u,x)=0$, and
$p$ in the relative interior of $\Delta_{\mathcal S}$,
\begin{equation}
\left.\frac{d}{d\varepsilon}F(p+\varepsilon h)
\right|_{\varepsilon=0}=\langle h,\Omega[p]\rangle.
\label{eq:score-gradient}
\end{equation}

\subsection{The relative-entropy identity}

\begin{proposition}[Score identity and relative smoothness]
\label{prop:score-identity}
Let $p,q$ be joint laws with $\supp p\subseteq\supp q$.
Their induced output laws satisfy
\begin{align}
\langle p,\Omega[p]-\Omega[q]\rangle
={}&D(p_{UZ}\|q_{UZ})-D(p_Z\|q_Z)\nonumber\\
&-D(p_{UY}\|q_{UY})+D(p_Y\|q_Y).
\label{eq:score-identity}
\end{align}
Consequently,
\begin{equation}
-D(p_{UY}\|q_{UY})
\le\langle p,\Omega[p]-\Omega[q]\rangle
\le D(p_{UZ}\|q_{UZ})\le D(p\|q),
\label{BN-classical-simple}
\end{equation}
and
\begin{equation}
\big|\langle p,\Omega[p]-\Omega[q]\rangle\big|
\le D(p\|q).
\label{BN}
\end{equation}
In particular, if $q$ is positive on $\mathcal S$ and
$p\in\Delta_{\mathcal S}$, then
\begin{equation}
\left|F(p)-F(q)-\langle p-q,\Omega[q]\rangle\right|
\le D(p\|q).
\label{eq:relative-smoothness}
\end{equation}
\end{proposition}

\begin{proof}
For either output $A$, expansion of the conditional logarithms gives
\[
\sum_{u,a}p_{UA}(u,a)
\log\frac{p_{A|u}(a)}{q_{A|u}(a)}
=D(p_{UA}\|q_{UA})-D(p_U\|q_U).
\]
Subtracting the $Y$ expression from the $Z$ expression cancels
$D(p_U\|q_U)$; the two unconditional logarithms supply the remaining
terms in \eqref{eq:score-identity}. The support assumption makes every
relative entropy here finite. Put
$\Delta_A=D(p_{UA}\|q_{UA})-D(p_A\|q_A)$.
Data processing gives $0\le\Delta_A\le D(p_{UA}\|q_{UA})\le D(p\|q)$.
The identity is $\Delta_Z-\Delta_Y$, proving the bounds.
Finally, \eqref{eq:F-as-Omega} identifies the left-hand side of
\eqref{eq:relative-smoothness} with that of \eqref{BN}.
\end{proof}

\subsection{The update and quantitative descent}

For $p$ positive on $\mathcal S$, set
\begin{equation}
\mathcal F[p](u,x)
=\frac{p(u,x)e^{-\Omega[p](u,x)}}{Z(p)},\qquad
Z(p)=\sum_{(v,a)\in\mathcal S}p(v,a)e^{-\Omega[p](v,a)},
\label{eq:F-map}
\end{equation}
and iterate
\begin{equation}
p^{(t+1)}=\mathcal F[p^{(t)}].
\label{eq:iteration}
\end{equation}
We write $c=Z(p)^{-1}$, so that
\begin{equation}
c^{-1}=\sum_{(u,x)\in\mathcal S}p(u,x)e^{-\Omega[p](u,x)}.
\label{eq:c-normalizer}
\end{equation}
Equivalently, the update is the entropic mirror-descent step
\begin{equation}
\mathcal F[p]
=\argmin_{r\in\Delta_{\mathcal S}}
\{\langle r,\Omega[p]\rangle+D(r\|p)\}.
\label{eq:mirror-descent}
\end{equation}
The relative-smoothness bound above supplies a unit step size without
a line search.

\begin{theorem}[Quantitative descent]
\label{thm:classical-descent}
For $q=\mathcal F[p]$,
\begin{equation}
F(q)\le F(p)-D(p\|q).
\label{eq:F-drop}
\end{equation}
Thus the objective decreases strictly unless the update leaves $p$
unchanged.
\end{theorem}

\begin{proof}
The update implies
\begin{equation}
\log\frac{q(u,x)}{p(u,x)}=\log c-\Omega[p](u,x).
\label{eq:log-update}
\end{equation}
Averaging first under $p$ and then under $q$ yields
\begin{equation}
D(p\|q)=F(p)-\log c,\qquad
D(q\|p)=\log c-\langle q,\Omega[p]\rangle.
\label{eq:KL-to-update}
\end{equation}
Apply \eqref{BN} to $(q,p)$ to obtain
\[
F(q)\le\langle q,\Omega[p]\rangle+D(q\|p)
      =\log c=F(p)-D(p\|q).
\]
\end{proof}

\begin{algorithm}[t]
\caption{Multiplicative optimization of the wiretap secrecy objective}
\label{alg:wiretap}
\textbf{Input:} Channels $W_Y,W_Z$; allowed support $\mathcal S$;
initial law $p^{(0)}>0$ on $\mathcal S$; maximum iteration count $T$.
\par\smallskip
\textbf{For} $t=0,\ldots,T-1$:
\begin{enumerate}
\item Compute $p_U^{(t)},p_{UY}^{(t)},p_{UZ}^{(t)},p_Y^{(t)},p_Z^{(t)}$.
\item Evaluate $\Omega[p^{(t)}](u,x)$ by \eqref{eq:Omega-def}
on $\mathcal S$.
\item Set $a_{ux}=\log p^{(t)}(u,x)-\Omega[p^{(t)}](u,x)$ and
$m=\max_{(u,x)\in\mathcal S}a_{ux}$.
\item Set $p^{(t+1)}(u,x)=
e^{a_{ux}-m}/\sum_{(v,b)\in\mathcal S}e^{a_{vb}-m}$.
\end{enumerate}
\textbf{Output:} $p^{(T)}$ and the achieved rate $-F(p^{(T)})/\log2$ in bits.
\end{algorithm}

The cost per iteration is
$O(|\mathcal U||\mathcal X|(|\mathcal Y|+|\mathcal Z|))$.
Logarithmic weights and the displayed shifted normalization avoid
overflow; implementations should also avoid truncating small positive
weights to zero. An objective increment or an iterate difference can
be used as a numerical stopping criterion, but neither is a global
optimality certificate in the nonconvex case. Independent restarts
produce additional feasible rates. Since choosing $U$ constant gives
rate zero, a reported achievable lower bound can always be replaced
by $\max\{0,-F(p^{(T)})/\log2\}$.

\subsection{Comparison with earlier wiretap algorithms}
\label{subsec:comparison-earlier-wiretap}

The wiretap update in the preprint version of~\cite{Haya}
(arXiv:2302.06905v3, Sec.~3.4.1, Eqs.~(87)--(88))
uses a score that depends on $u$ but is constant across
the input symbols $x$ in each $u$-row.
In the notation of this paper, the score in Eq.~(87) of that preprint
is the average of \eqref{eq:Omega-def} over the conditional law
$p_{X|U=u}$:
\begin{equation}
\sum_x p_{X|U}(x|u)\Omega[p](u,x)
= D(p_{Z|u}\|p_Z)-D(p_{Y|u}\|p_Y).
\label{eq:earlier-wiretap-score}
\end{equation}
Here $p_U(u)>0$ is understood.
As the multiplicative factor in that update does not
depend on $x$, the conditional law $p_{X|U=u}$
is unchanged. By contrast, our update~\eqref{eq:F-map}
gives
\begin{equation}
\mathcal F[p](x|u)=
\frac{p_{X|U}(x|u)e^{-\Omega[p](u,x)}}
{\sum_a p_{X|U}(a|u)e^{-\Omega[p](u,a)}}.
\label{eq:updated-prefixing-law}
\end{equation}
The update therefore changes the conditional
encoder whenever \eqref{eq:Omega-def} varies across
the occupied input symbols $x$ in that row.

In that preprint, the sufficient convergence conditions were not
verified for the general wiretap formulation; they were verified for a
separate degraded-channel formulation. For a fixed prefix, the unit-step
row update also admits a descent bound: apply the relative-entropy
argument to the effective channels $U\to Y$ and $U\to Z$, treating $U$
as the channel input.

Our update~\eqref{eq:F-map} jointly updates $p_U$ and $p_{X|U}$.
The distinction is a quantitative descent guarantee while the
stochastic prefix is optimized together with the auxiliary law.
The objective decreases by at least $D(p^{(t)}\|p^{(t+1)})$ at each
iteration, as shown in~\eqref{eq:F-drop}, and strictly unless the
iterate is fixed. This result holds for arbitrary finite classical
wiretap channels, without assuming degradedness or less noisiness.

Although our update differs from the wiretap update in~\cite{Haya}, it
is a special case of the general Algorithm~1 of that preprint, obtained
by taking the mixture family $\mathcal M=\mathcal P(\mathcal U\times\mathcal X)$,
$\Psi[p]=\Omega[p]$, and $\gamma=1$.

\subsection{What converges for arbitrary finite channels}

\begin{theorem}[Values, increments, and boundary accumulation points]
\label{thm:classical-limit}
For any finite channel pair and any initial law positive on
$\mathcal S$, there is a finite $F_\infty$ such that
\begin{equation}
F(p^{(t)})\downarrow F_\infty,
\qquad
\sum_{t=0}^{\infty}D(p^{(t)}\|p^{(t+1)})
\le F(p^{(0)})-F_\infty.
\label{eq:monotone-F}
\end{equation}
In particular,
$\|p^{(t+1)}-p^{(t)}\|_1\to0$.
Every accumulation point $p^*$ satisfies
\begin{equation}
F(p^*)=F_\infty,\qquad
\Omega[p^*](u,x)=F_\infty
\quad\text{whenever }p^*(u,x)>0.
\label{eq:face-stationarity}
\end{equation}
Thus every accumulation point is stationary relative to its support
face, including when some coordinates or entire $U$-rows vanish.
An accumulation point positive on all of $\mathcal S$ is an ordinary
stationary point of the optimization over $\Delta_{\mathcal S}$.
\end{theorem}

\begin{proof}
The objective is continuous on the compact simplex, so it is bounded
below. Theorem~\ref{thm:classical-descent} and telescoping prove
\eqref{eq:monotone-F}; Pinsker's inequality gives the iterate assertion.
Take a subsequence $p^{(t_k)}\to p^*$ and an occupied coordinate
$i=(u,x)$ of $p^*$. Asymptotic regularity implies
$p^{(t_k+1)}(i)/p^{(t_k)}(i)\to1$.
Every channel output with $W_A(a|x)>0$ has
$p^*_{UA}(u,a)\ge p^*(u,x)W_A(a|x)>0$.
Consequently the score at this coordinate is continuous along the
subsequence, even if other output probabilities vanish.
Furthermore, \eqref{eq:KL-to-update} gives
\[
\log c_t=F(p^{(t)})-D(p^{(t)}\|p^{(t+1)})\longrightarrow F_\infty.
\]
Taking limits in \eqref{eq:log-update} proves
\eqref{eq:face-stationarity}. Equation~\eqref{eq:score-gradient},
applied on the support face of $p^*$, proves the stationarity claim.
\end{proof}

At a differentiable boundary point, full-simplex first-order
optimality additionally requires $\Omega[p^*](u,x)\ge F(p^*)$ at
unoccupied allowed coordinates. The next theorem establishes
convergence of the entire sequence for strictly positive channel
matrices, including boundary limits.

\subsection{Convergence of the entire sequence for positive channels}

\begin{theorem}[Finite length in square-root coordinates]
\label{thm:positive-channel-convergence}
Suppose $W_Y(y|x)>0$ and $W_Z(z|x)>0$ for every input and output
symbol, and initialize positively on $\mathcal S$. Then
\begin{equation}
\sum_{t=0}^{\infty}
\left\|\sqrt{p^{(t+1)}}-\sqrt{p^{(t)}}\right\|_2<\infty.
\label{eq:sqrt-finite-length}
\end{equation}
In particular, $p^{(t)}$ converges to one distribution $p^*$ satisfying
\eqref{eq:face-stationarity}.
For every surviving row $p_U^*(u)>0$, its unoccupied allowed
coordinates also satisfy
\begin{equation}
\Omega[p^*](u,x)\ge F(p^*).
\label{eq:surviving-row-KKT}
\end{equation}
If every allowed $U$-row survives, the limit satisfies the full
first-order Karush--Kuhn--Tucker conditions on
$\Delta_{\mathcal S}$.
\end{theorem}

\begin{proof}
Let $\delta_A=\min_{a,x}W_A(a|x)>0$ for $A=Y,Z$.
Both conditional and unconditional output probabilities lie in
$[\delta_A,1]$, so
\begin{equation}
|\Omega[p](u,x)|\le B:=\log(1/\delta_Y)+\log(1/\delta_Z)
\label{eq:bounded-scores}
\end{equation}
at every occupied coordinate.

Put $s=\sqrt p$ coordinatewise, so $\|s\|_2=1$, and let
$G(s)=F(s^2)$ on the sphere in $\mathbb R^{|\mathcal S|}$.
For differentiation one can extend $F$ homogeneously off the
simplex using \eqref{eq:F-entropy-form}.
The function $G$ is continuously differentiable even at points
where an entire $U$-row vanishes. Indeed, if
$w_u=\sum_xs_{ux}^2$, the row terms can be written as
$w_u h_u((s_{ux}^2)_x/w_u)$, with $h_u$ smooth with bounded
derivatives on the closed conditional-input simplex. A vanishing
row therefore contributes $O(w_u)$ to $G$ and
$O(\sqrt{w_u})$ to its gradient; extension by zero is $C^1$.
The unconditional terms are smooth by channel positivity.
Moreover, $G$ restricted to the sphere is globally subanalytic:
its formulas use algebraic operations and logarithms with arguments
in fixed compact subsets of $(0,\infty)$, where the logarithm is
restricted analytic, together with the specified zero-row
extensions. The subanalytic \L ojasiewicz gradient inequality
therefore applies on the sphere~\cite{BolteKL2007}.

Write $s_t=\sqrt{p^{(t)}}$, $d_t=\|s_{t+1}-s_t\|_2$, and
$P_s=I-ss^{\mathsf T}$ for tangent projection. At the iterates,
\begin{equation}
\operatorname{grad}_{\mathbb S}G(s_t)
=2P_{s_t}(s_t\mathbin{\odot}\Omega[p^{(t)}]),
\label{eq:sphere-gradient}
\end{equation}
where $\odot$ denotes coordinatewise multiplication. If
$a_i=(\Omega[p^{(t)}](i)-\log c_t)/2$, then
$s_{t+1,i}=s_{t,i}e^{-a_i}$. Since
$|\log c_t|\le B$, we have $|a_i|\le B$. The elementary inequality
$|e^{-a}-1|\ge e^{-B}|a|$ for $|a|\le B$, together with
$P_{s_t}s_t=0$, yields
\begin{equation}
\|\operatorname{grad}_{\mathbb S}G(s_t)\|_2
\le4e^B d_t.
\label{eq:gradient-step-control}
\end{equation}
Relative entropy dominates squared Hellinger distance, so descent
also gives
\begin{equation}
G(s_t)-G(s_{t+1})
\ge D(p^{(t)}\|p^{(t+1)})\ge d_t^2.
\label{eq:square-root-descent}
\end{equation}

Let $\ell=F_\infty$. The accumulation set of $s_t$ is compact,
$G$ is constant there with value $\ell$, and
\eqref{eq:gradient-step-control} makes its points critical.
The \L ojasiewicz inequality, applied in finitely many sphere
charts covering this set, supplies a concave increasing function
$\varphi(v)=c_0v^{1-\theta}$, with $0\le\theta<1$, such that
\[
\varphi'(G(s_t)-\ell)
\|\operatorname{grad}_{\mathbb S}G(s_t)\|_2\ge1
\]
for all sufficiently large $t$ with $G(s_t)>\ell$.
Concavity of $\varphi$ and the last two bounds imply
\begin{align*}
\varphi(G(s_t)-\ell)-\varphi(G(s_{t+1})-\ell)
&\ge\varphi'(G(s_t)-\ell)d_t^2\\
&\ge d_t/(4e^B).
\end{align*}
Telescoping proves \eqref{eq:sqrt-finite-length}. If the objective
reaches $\ell$ at a finite iteration, descent makes every later
step zero instead. This proves convergence of $s_t$ and $p^{(t)}$.

Finally, at an unoccupied coordinate in a surviving row, strict
channel positivity makes $\Omega[p^{(t)}](u,x)$ converge to its
finite limiting score. If that score were below $F(p^*)$, then
\eqref{eq:log-update} would make the positive mass at this coordinate
increase by a factor larger than $1+\varepsilon$ at every sufficiently
late iteration, contradicting convergence of that mass to zero.
This proves \eqref{eq:surviving-row-KKT}.
\end{proof}

Channels with zero transition probabilities retain the guarantees of
Theorem~\ref{thm:classical-limit}. For vanished auxiliary rows, the
conclusion remains support-face stationarity.

\begin{example}[A full-support fixed point need not be a local minimum]
\label{ex:independent-fixed-point}
Let Bob receive $X$ through a binary symmetric channel with crossover
$0<a<1/2$, and let Eve receive a constant. Set
$p(u,x)=1/4$ for binary $U,X$. Independence gives
$\Omega[p](u,x)=0$ everywhere, so the iteration remains at this
full-support law and $F(p)=0$. Replacing its diagonal entries by
$1/4+\varepsilon$ and its off-diagonal entries by
$1/4-\varepsilon$, for arbitrarily small nonzero $\varepsilon$,
makes $I(U;Y)>0$ and hence $F<0$.
Thus even with strictly positive channel matrices, convergence to
a local minimum cannot be guaranteed for every initialization.
\end{example}

\subsection{Global convergence in the restricted less-noisy regime}

\begin{theorem}[Convex objective: global convergence and a value bound]
\label{thm:convex-convergence}
Suppose $F$ is convex on $\Delta_{\mathcal S}$ and
$p^{(0)}>0$ on $\mathcal S$. The entire sequence converges to a
global minimizer. For every global minimizer $p^*$ and every
$T\ge1$,
\begin{equation}
0\le F(p^{(T)})-F(p^*)\le\frac{D(p^*\|p^{(0)})}{T}.
\label{eq:convex-rate}
\end{equation}
For a less-noisy classical wiretap channel, this applies to the
diagonal support $U=X$, yielding convergence to a
capacity-achieving input distribution. With a uniform initial
input, the capacity gap in bits is at most
$\log|\mathcal X|/(T\log2)$.
\end{theorem}

\begin{proof}
Let $p=p^{(t)}$, $q=p^{(t+1)}$, and let $r$ be any minimizer.
Convexity, \eqref{eq:score-gradient}, and
\eqref{eq:F-as-Omega} imply
$F(r)\ge F(p)+\langle r-p,\Omega[p]\rangle
=\langle r,\Omega[p]\rangle$.
Meanwhile, \eqref{eq:log-update} gives
\[
D(r\|p)-D(r\|q)=\log c-\langle r,\Omega[p]\rangle.
\]
Using $F(q)\le\log c$ from the descent proof therefore yields
\begin{equation}
0\le F(q)-F(r)
\le D(r\|p)-D(r\|q).
\label{eq:convex-telescope}
\end{equation}
Summing and using monotonicity proves \eqref{eq:convex-rate}.
Compactness gives an accumulation point $\bar p$, and the value
bound makes it optimal. The quantities
$D(\bar p\|p^{(t)})$ are nonincreasing by
\eqref{eq:convex-telescope}; along a subsequence converging to
$\bar p$ they tend to zero, including when $\bar p$ has zero
coordinates. Hence they tend to zero along the entire sequence,
and Pinsker's inequality proves $p^{(t)}\to\bar p$.
For less-noisy channels, convexity of
$F(p_X)=I(X;Z)-I(X;Y)$ and sufficiency of $U=X$ were established
in the preceding discussion. Finally,
$D(p^*\|p^{(0)})\le\log|\mathcal X|$ for uniform $p^{(0)}$.
\end{proof}

\section{Classical numerical experiments}
\label{sec:classical-numerics}

We evaluate three distinct features of the iteration: agreement with
known capacities, dependence on initialization outside the concave
regime, and optimization of a genuinely useful stochastic prefix.
All rates in this section are in bits per channel use.

\subsection{Implementation and numerical benchmarks}
\label{subsec:numerical-protocol}

The random seed is $20261011$, and random initial laws are sampled from
$\operatorname{Dirichlet}(1,1,1,1)$ on the $2\times2$ joint simplex.
Where indicated, we also use nine diagonal initial laws with biases
$0.05,0.1625,\ldots,0.95$, each mixed with weight $10^{-4}$ of the
uniform joint law.

The stopping test requires both a maximum entrywise change at most
$10^{-11}$ and the first-order residual
\begin{equation}
  \mathcal G(p):=\langle p,\Omega[p]\rangle-
                 \min_{u,x}\Omega[p](u,x)\le 10^{-7}
  \quad\text{(nats)},
  \label{eq:numerical-gap}
\end{equation}
with a cap of $60000$ iterations. This is a first-order stationarity
test, not a global optimality certificate for a nonconvex objective.
The implementation records any decreases in the independently
evaluated objective, so monotonicity can also be checked numerically.
Updates are evaluated in the log domain; a numerical probability
floor of $10^{-300}$ prevents underflow.
The channel probabilities printed below are normalized by
column after rounding; this affects only their last displayed digits.

For binary input define
\[
 f(p):=I(X;Y)-I(X;Z),\qquad p=P_X(0).
\]
The identity
\begin{equation}
 I(U;Y)-I(U;Z)
 = f\!\left(\sum_u p_U(u)p_u\right)
   -\sum_u p_U(u)f(p_u),\qquad p_u=P(X=0\mid U=u),
 \label{eq:binary-envelope-rate}
\end{equation}
provides a useful independent benchmark. Writing
$\operatorname{co}f$ for the lower convex envelope on $[0,1]$ gives
\begin{equation}
 C_s=\max_{0\le p\le1}\{f(p)-(\operatorname{co}f)(p)\}.
 \label{eq:binary-envelope-capacity}
\end{equation}
Indeed, at a fixed mean $p$, minimizing the second term of
\eqref{eq:binary-envelope-rate} is precisely the convex-envelope
problem, and in one dimension two support points suffice.

We sample $f$ on uniform meshes of widths $2\times10^{-4}$ and
$5\times10^{-5}$, compute their lower convex hulls, and maximize the
vertical gap at the mesh points. Each resulting rate is feasible:
its two conditional biases and mixing weights are explicitly
reconstructed. These are mesh approximations from below, rather than
certified continuum optima. The comparison of the two resolutions is
reported alongside the algorithmic values. This reduction is more
economical than gridding the entire joint simplex and supplies an
independent check on the joint implementation.

For the nonconcave examples we additionally run SciPy's SLSQP on the
same $24$ random initial joint laws, using the analytic score as the
gradient, a normalization equality, a numerical probability floor
$10^{-12}$, tolerance $10^{-12}$, and at most $2000$ iterations.
This provides an independent numerical cross-check.

\subsection{Known-capacity and more-capable benchmarks}
\label{subsec:known-capacity-benchmarks}

\begin{figure}[t]
 \centering
 \includegraphics[width=.94\linewidth]{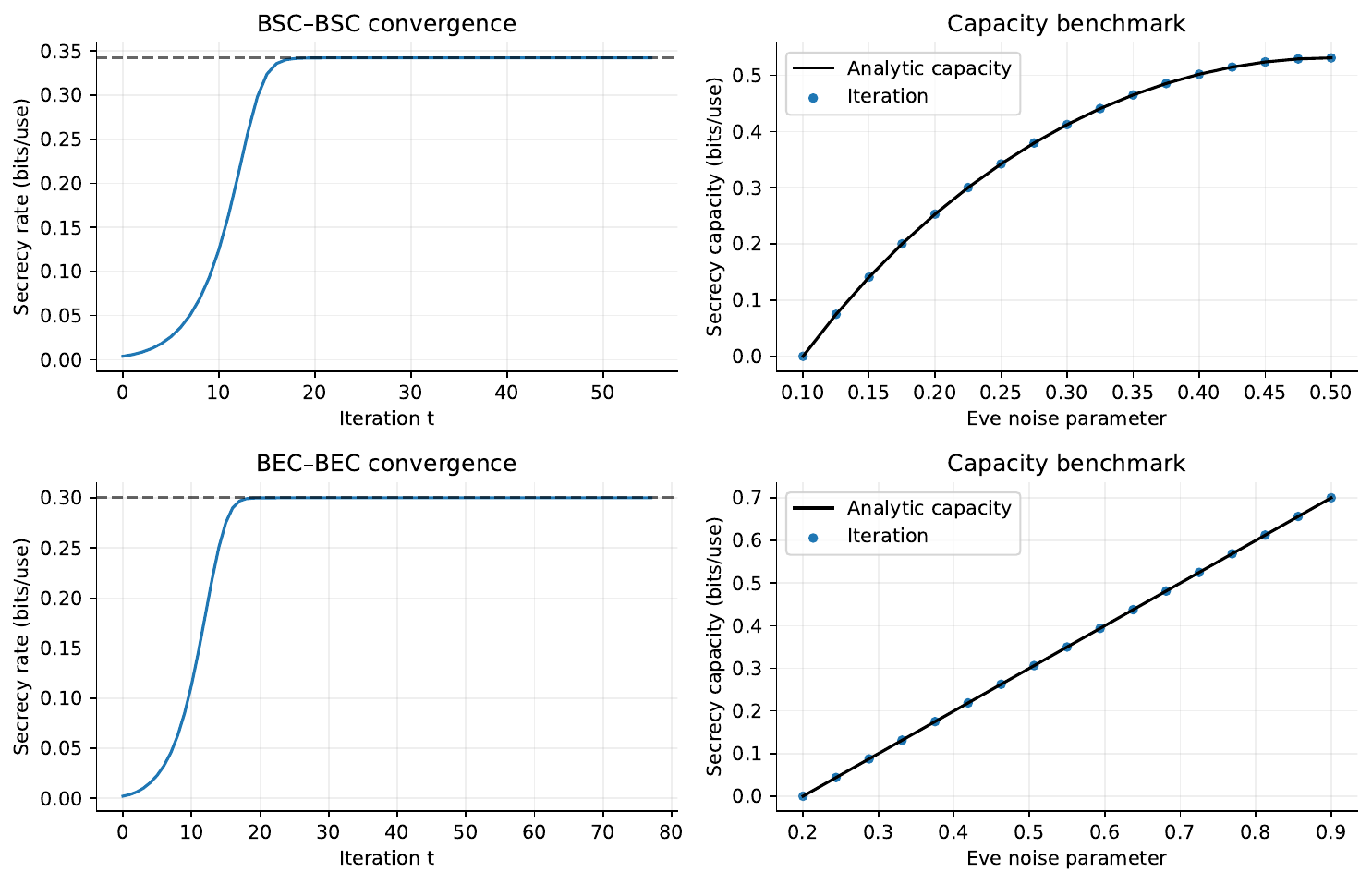}
 \caption{Known-capacity checks. Top: BSC--BSC with Bob crossover
 $p_1=0.1$; the trajectory uses Eve crossover $p_2=0.25$.
 Bottom: BEC--BEC with Bob erasure $\epsilon_1=0.2$; the trajectory
 uses Eve erasure $\epsilon_2=0.5$. The sweep markers agree with the
 analytic capacities at the displayed precision.}
 \label{fig:composite-known}
\end{figure}

For a BSC--BSC pair with $0\le p_1\le p_2\le1/2$, the secrecy capacity
is $h_2(p_2)-h_2(p_1)$. For a BEC--BEC pair with
$0\le\epsilon_1\le\epsilon_2\le1$, it is
$\epsilon_2-\epsilon_1$. Both are attained by a uniform input with
$U=X$. Figure~\ref{fig:composite-known} compares the joint iteration
with these formulas, using $p_1=0.1$ and $\epsilon_1=0.2$.
The individual trajectories start from one random full-support law;
each parameter-sweep marker is the best of four random starts and
the nine perturbed diagonal starts specified above.

\begin{figure}[t]
 \centering
 \includegraphics[width=.94\linewidth]{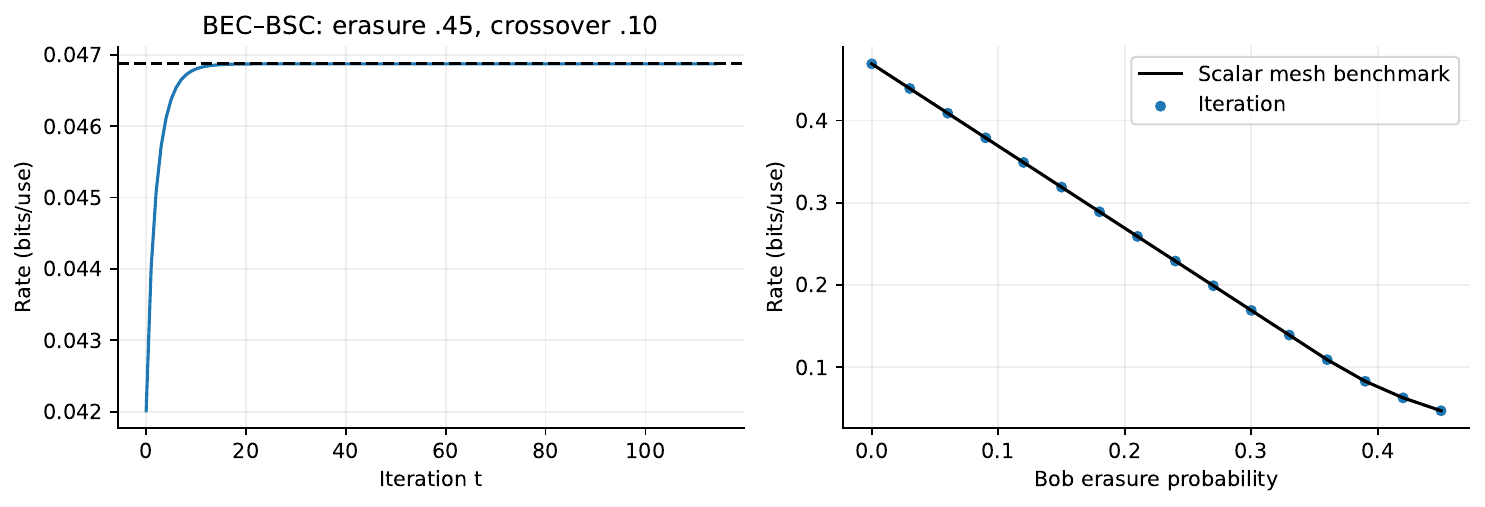}
 \caption{BEC--BSC validation with Eve crossover $\epsilon=0.1$.
 Left: a selected trajectory for Bob erasure $\alpha=0.45$.
 Right: comparison with the scalar mesh benchmark over
 $0\le\alpha\le0.45$. The more-capable condition makes $U=X$
 sufficient throughout this sweep.}
 \label{fig:composite-bec-bsc}
\end{figure}

We next take Bob's channel to be BEC$(\alpha)$ and Eve's channel
to be BSC$(\epsilon)$, fixing $\epsilon=0.1$.
For $\alpha<h_2(\epsilon)$ Bob is more capable, so prefixing is
unnecessary. The objective in the input bias is concave for
$\alpha\le4\epsilon(1-\epsilon)$; above that threshold it can be
nonconcave, although the capacity is still $\max_p f(p)$
\cite{OzelUlukusMoreCapable,ozel_ulukus_isit11}.
Here
\begin{equation}
 f(p)=(1-\alpha)h_2(p)
       -h_2\!\left(\epsilon+(1-2\epsilon)p\right)+h_2(\epsilon).
 \label{eq:Cs-bec-bsc}
\end{equation}
At $(\alpha,\epsilon)=(0.45,0.10)$ the scalar optimum is approximately
$0.04687458$ bits/use, at biases $0.09748$ and $0.90252$.
The uniform-input rate is only $0.01899559$ bits/use.
Figure~\ref{fig:composite-bec-bsc} shows the comparison.
The sweep uses eight random and nine perturbed diagonal starts at each
point; the displayed trajectory is selected from $24$ random and the
same nine diagonal starts. Agreement here tests the nonconcave input
optimization.

\subsection{Nonconcavity and dependence on initialization}
\label{subsec:num-bruteforce}

\begin{figure}[t]
 \centering
 \includegraphics[width=.94\linewidth]{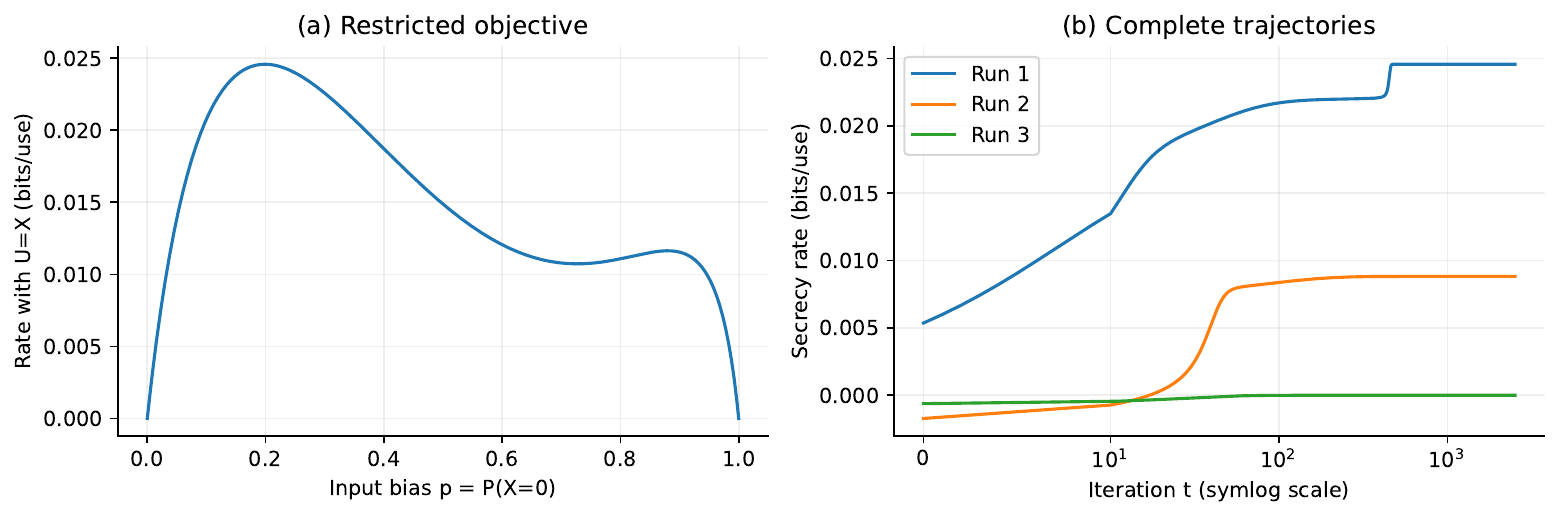}
 \caption{Channel 1. Left: the nonconcave objective with $U=X$.
 Right: three initializations of the joint iteration,
 continued through the numerical stopping test. The iteration axis
 is linear close to zero and logarithmic thereafter, so both the
 early transient and later approach are visible. The lower positive
 trajectory approaches an off-diagonal law.}
 \label{fig:nonconcave-case}
\end{figure}

Consider the following binary-input channels, whose columns are
indexed by $x=0,1$:
\begin{align}
 W^{(1)}_{Y|X}&=\begin{pmatrix}
 .01280667&.52296472\\ .48562477&.03454652\\ .50156856&.44248876
 \end{pmatrix},&
 W^{(1)}_{Z|X}&=\begin{pmatrix}
 .82518319&.20482578\\ .12881715&.12532976\\ .04599966&.66984446
 \end{pmatrix},\label{eq:chan1}\\
 W^{(2)}_{Y|X}&=\begin{pmatrix}
 .03781192&.46801438\\ .38646163&.25769582\\
 .21723845&.23980992\\ .35848801&.03447989
 \end{pmatrix},&
 W^{(2)}_{Z|X}&=\begin{pmatrix}
 .18544348&.64267425\\ .34532969&.30460267\\ .46922683&.05272307
 \end{pmatrix}.\label{eq:chan2}
\end{align}
The first channel has two restricted local maxima of $f(p)$,
approximately $0.02457978$ near $p=0.19901$ and $0.01163756$
near $p=0.8798$. The full joint iteration has additional behavior:
our random starts reach rates near $0$, $0.00882487$, and
$0.02457978$. A lower positive limiting law is approximately
\begin{equation}
 p_{UX}=\begin{pmatrix}
 .829945263&0\\ .092100696&.077954041
 \end{pmatrix}.
 \label{eq:nonconcave-lower-law}
\end{equation}
It is not in the diagonal family. The three trajectories in Fig.~\ref{fig:nonconcave-case}
are continued until the stated numerical stopping test.

\begin{figure}[t]
 \centering
 \includegraphics[width=.76\linewidth]{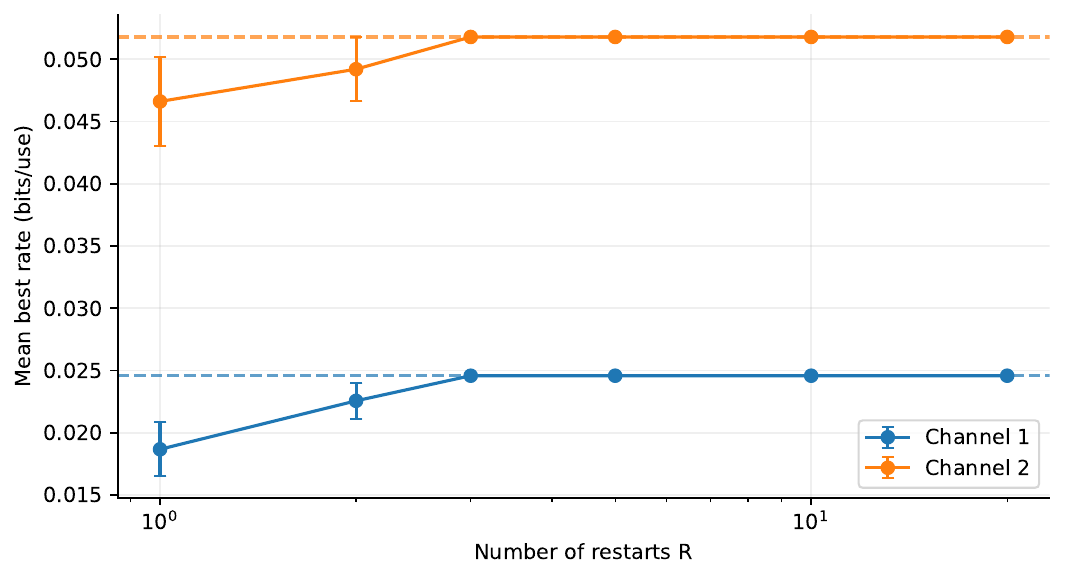}
 \caption{Mean best-of-$R$ terminal rate, with one-standard-error
 bars across $20$ independent trials of nested random starts.
 Dashed lines are the finer convex-envelope mesh benchmarks.
 Mesh values approximate the global optimum from below.}
 \label{fig:bestofR}
\end{figure}

For each of the two channels we run $20$ independent trials, each
containing a nested pool of $20$ random starts. For
$R\in\{1,2,3,5,10,20\}$ we record the best terminal rate among the
first $R$ starts. Figure~\ref{fig:bestofR} reports the mean and its
standard error across the $20$ trials. The nesting ensures that each
trial's best value is nondecreasing in $R$.
All $400$ runs per channel satisfy the numerical stopping test.
Restarts improve the attained rates in these examples. The best
observed rates also agree with diagonal solutions, illustrating
initialization dependence.

\subsection{A strict classical prefixing gain}
\label{subsec:classical-prefixing}

\begin{figure}[t]
 \centering
 \includegraphics[width=.94\linewidth]{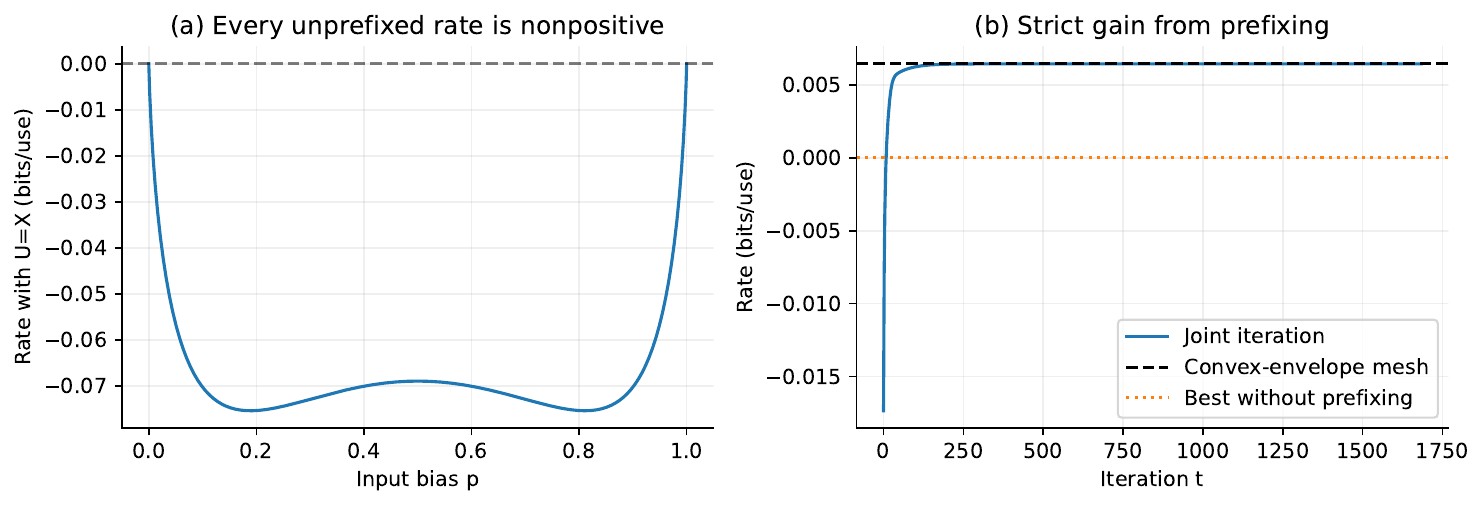}
 \caption{Bob BSC$(0.1)$, Eve BEC$(0.4)$. Left: every rate with
 $U=X$ is nonpositive. Right: the joint iteration finds a positive
 rate of approximately $0.00645039$ bits/use, consistent with the
 finer convex-envelope mesh benchmark.}
 \label{fig:classical-prefixing}
\end{figure}

To test the additional freedom supplied by $p_{UX}$, reverse the
mixed pair: let Bob observe BSC$(0.1)$ and Eve BEC$(0.4)$.
Now
\begin{equation}
 f(p)=h_2(0.1+0.8p)-h_2(0.1)-0.6h_2(p).
\end{equation}
Eve is more capable because $0.4<h_2(0.1)$, so $f(p)\le0$ and
the best rate with $U=X$ is zero. Nevertheless, a stochastic prefix
produces a positive secrecy rate. The joint iteration and the
convex-envelope benchmark give approximately
$0.00645039$ bits/use. One corresponding law has uniform $U$ and
\[
 P(X=0\mid U=0)\simeq0.1892636,\qquad
 P(X=0\mid U=1)\simeq0.8107364.
\]
Thus the example tests the joint optimization in a regime where
optimizing the input bias alone cannot achieve a positive rate.
Figure~\ref{fig:classical-prefixing} shows the trajectory and
the no-prefixing comparison. The run uses $24$ random and nine
perturbed diagonal starts.

\begin{table}[t]
\centering\small
\caption{Numerical cross-checks in bits/use. The two mesh columns are feasible convex-envelope approximations. The multiplicative update and SLSQP columns give the best result from the same 24 random starts, without diagonal warm-starts.}
\label{tab:numerical-crosscheck}
\begin{tabular}{lrrrr}
\hline
Channel & Mesh $2\cdot10^{-4}$ & Mesh $5\cdot10^{-5}$ & Update & SLSQP \\
\hline
Channel 1 & 0.0245797751 & 0.0245797751 & 0.0245797752 & 0.0245797752 \\
Channel 2 & 0.0517913891 & 0.0517913919 & 0.0517913919 & 0.0517913919 \\
BSC--BEC prefix & 0.0064503851 & 0.0064503865 & 0.0064503866 & 0.0064503866 \\
\hline
\end{tabular}
\end{table}

The SLSQP comparison in Table~\ref{tab:numerical-crosscheck} uses
the same $24$ random initializations as the corresponding subset of
the multiplicative runs. Both methods find matching high rates here.
The multiplicative method combines an explicit update with a general
descent inequality.
\FloatBarrier

\section{Classical--quantum extension and covariant preprocessing}
\label{sec:quantum-single-letter}

We extend the construction to the single-letter private information
of finite-input classical--quantum (cq) wiretap channels, including
noncommuting output states.
We first establish the quantum score identity and descent bound, then
derive a structured version of the update for the covariant
preprocessing family of Hayashi and Salek~\cite{SalekWiretap}.

\subsection{Functional, supports, and quantum score}

Let $x\in\cX$ produce states $\rho_x^B\in\cD(\cH_B)$ and
$\sigma_x^E\in\cD(\cH_E)$, with finite-dimensional output spaces.
Only these two marginals enter the private-information objective, so
we may represent the induced state using the product coupling
\begin{equation}
 \rho_{UXBE}^{(p)}=\sum_{u,x}p(u,x)|u\rangle\!\langle u|
 \otimes|x\rangle\!\langle x|\otimes\rho_x^B\otimes\sigma_x^E.
 \label{eq:cq-state}
\end{equation}
Write $\rho_B^p=\sum_{u,x}p(u,x)\rho_x^B$ and
$\rho_{B|u}^p=\sum_xp(x|u)\rho_x^B$, with analogous notation
$\sigma_E^p,\sigma_{E|u}^p$ for Eve. Rows with $p_U(u)=0$ are
omitted from conditional averages. The objective is the continuous
functional
\begin{equation}
 F_Q(p):=I(U;E)_p-I(U;B)_p,
 \qquad P^{(1)}(W)=-\min_p F_Q(p).
 \label{eq:FQ-def}
\end{equation}
One may take $|\cU|\leq|\cX|$ in this minimization
\cite[Lemma~2]{TikkuBertaRenes}; the relation to operational capacity
is given in Section~\ref{subsec:ordering}.

For strictly positive $p$,
restrict Bob's space to $\operatorname{supp}\sum_x\rho_x^B$ and
Eve's to $\operatorname{supp}\sum_x\sigma_x^E$. The conditional and
average states are positive definite on these respective spaces,
even when the individual channel outputs are singular. Define
\begin{align}
 \Omega_Q[p](u,x)
 &:=\Tr\,\sigma_x^E(\log\sigma_{E|u}^p-\log\sigma_E^p)
 \nonumber\\[-1mm]
 &\quad-\Tr\,\rho_x^B(\log\rho_{B|u}^p-\log\rho_B^p).
 \label{eq:OmegaQ-def}
\end{align}
At a boundary distribution, this score is needed only for
$p(u,x)>0$; the traces are then taken on the supports of the states
inside the logarithms. Thus no expression of the form
$0\cdot\infty$ is required. Direct expansion gives
\begin{equation}
 F_Q(p)=\sum_{u,x:p(u,x)>0}p(u,x)\Omega_Q[p](u,x).
 \label{eq:FQ-as-OmegaQ}
\end{equation}
Moreover, on the interior of the simplex, $\Omega_Q[p]$ is a gradient
of $F_Q$ along zero-sum perturbations. This follows by differentiating
$S(A)=-\Tr\,A\log A$ and cancelling the trace terms.

\subsection{Relative-entropy identity and descent}

For a receiver $R=B,E$, denote the corresponding output marginal by
$\tau_R^p$ and the classical--quantum joint marginal by $\tau_{UR}^p$;
thus $\tau_B^p=\rho_B^p$ and $\tau_E^p=\sigma_E^p$. Define
\[
 \Gamma_R(p,q):=D(\tau_{UR}^p\Vert\tau_{UR}^q)
                    -D(\tau_R^p\Vert\tau_R^q).
\]
Here $D$ is Umegaki relative entropy for states and ordinary relative
entropy for probability distributions.

\begin{proposition}[Quantum score identity]
\label{prop:quantum-score}
For strictly positive $p,q$,
\begin{equation}
 \langle p,\Omega_Q[p]-\Omega_Q[q]\rangle
       =\Gamma_E(p,q)-\Gamma_B(p,q),
 \label{eq:score-id-Q}
\end{equation}
and
\begin{equation}
 \big|\langle p,\Omega_Q[p]-\Omega_Q[q]\rangle\big|
       \leq D(p\Vert q).
 \label{eq:BN-Q}
\end{equation}
The identities also hold for $\operatorname{supp}p\subseteq
\operatorname{supp}q$, with scores evaluated only on
$\operatorname{supp}p$.
\end{proposition}

\begin{proof}
The block-diagonal relative-entropy identity gives
\[
 D(\tau_{UR}^p\Vert\tau_{UR}^q)
 =D(p_U\Vert q_U)+\sum_up_U(u)
                         D(\tau_{R|u}^p\Vert\tau_{R|u}^q).
\]
Substituting~\eqref{eq:OmegaQ-def} and cancelling
$D(p_U\Vert q_U)$ proves~\eqref{eq:score-id-Q}.
Tracing out $U$ gives $\Gamma_R(p,q)\geq0$. The cq preparation
map $(U,X)\mapsto(U,R)$ gives
$\Gamma_R(p,q)\leq D(\tau_{UR}^p\Vert\tau_{UR}^q)
\leq D(p\Vert q)$ by data processing. The difference of two numbers
in $[0,D(p\Vert q)]$ satisfies~\eqref{eq:BN-Q}.
\end{proof}

The multiplicative step is
\begin{equation}
 q(u,x)={\cal F}_Q[p](u,x)
 =\frac{p(u,x)e^{-\Omega_Q[p](u,x)}}{Z_Q(p)},
 \qquad Z_Q(p)=\sum_{v,z}p(v,z)e^{-\Omega_Q[p](v,z)}.
 \label{eq:FQ-map}
\end{equation}
It preserves strict positivity.

\begin{proposition}[Quantum descent and limit properties]
\label{prop:quantum-descent}
For a strictly positive starting law, the iterates of
\eqref{eq:FQ-map} satisfy
\begin{equation}
 F_Q(p^{(t+1)})\leq F_Q(p^{(t)})
                    -D(p^{(t)}\Vert p^{(t+1)}).
 \label{eq:monotone-FQ}
\end{equation}
Their objective values converge, and
\[
 \sum_{t=0}^{\infty}D(p^{(t)}\Vert p^{(t+1)})<\infty,
 \qquad \|p^{(t+1)}-p^{(t)}\|_1\longrightarrow0.
\]
Every accumulation point $p^*$ is stationary on its support face:
$\Omega_Q[p^*](u,x)=F_Q(p^*)$ whenever $p^*(u,x)>0$.
\end{proposition}

\begin{proof}
Apply~\eqref{eq:BN-Q} to $(q,p)$, where $q={\cal F}_Q[p]$.
The exponential update then gives
\begin{align*}
 F_Q(q)&\leq\langle q,\Omega_Q[p]\rangle+D(q\Vert p)
       =-\log Z_Q(p)\\
       &=F_Q(p)-D(p\Vert q).
\end{align*}
The lower boundedness and continuity of $F_Q$ on the finite simplex,
telescoping, and Pinsker's inequality give the first claims, as in
Theorem~\ref{thm:classical-limit}.

For completeness, scores on positive coordinates are continuous
along a sequence tending to a boundary point. For such a coordinate,
each conditional or marginal output $A$ in~\eqref{eq:OmegaQ-def}
satisfies $A\geq c\rho_x^B$ or $A\geq c\sigma_x^E$ with a fixed
$c>0$ in a neighborhood of the limit. If $\lambda$ is a small
eigenvalue of $A$ with spectral projection $\Pi_\lambda$, then
$\Tr(\rho_x^B\Pi_\lambda)\leq\lambda
\Tr\Pi_\lambda/c$, and similarly for Eve.
Consequently the small-eigenvalue contributions to the trace of the
logarithm vanish with $\lambda|\log\lambda|$; the remaining part is
continuous. Along a convergent subsequence, asymptotic regularity
makes $p^{(t+1)}(u,x)/p^{(t)}(u,x)\to1$ on positive limit
coordinates. Taking logarithms in~\eqref{eq:FQ-map}, and using
$-\log Z_Q(p^{(t)})=F_Q(p^{(t)})-
D(p^{(t)}\Vert p^{(t+1)})$, proves the score equality.
\end{proof}

The algorithm can therefore be applied to any fixed block channel
$W^{\otimes n}$ by replacing $x$ by $x^n$. An obtained value divided
by $n$ is an achievable private rate. The block alphabets and output
dimensions grow exponentially with $n$.

\subsection{Monotone optimization of covariant preprocessing}
\label{subsec:covariant-updates}

Hayashi and Salek~\cite[Theorem~IV.1]{SalekWiretap} consider
group-orbit cq wiretap channels and a useful family of randomized
encoders. Let the input alphabet be a finite group $G$ and suppose
$\rho_{gx}^B=V_g^B\rho_x^B(V_g^B)^\dagger$ and
$\sigma_{gx}^E=V_g^E\sigma_x^E(V_g^E)^\dagger$ for unitary
representations (projective representations also suffice).
For independent $V\sim a$ and $Z\sim b$, the encoder sends $X=VZ$:
\begin{equation}
 p_{a,b}(v,x)=a(v)b(v^{-1}x),\qquad
 (a*b)(x)=\sum_va(v)b(v^{-1}x).
 \label{eq:covariant-joint}
\end{equation}
Their notation is $a=P_3$, $b=P_1$. With
\[
 \Delta(c):=S\!\left(\sum_xc(x)\rho_x^B\right)
                  -S\!\left(\sum_xc(x)\sigma_x^E\right),
\]
covariance makes each receiver's conditional entropy independent of
$v$, and hence their rate formula is
\begin{equation}
 R(a,b):=I(V;B)-I(V;E)
        =\Delta(a*b)-\Delta(b)=-F_Q(p_{a,b}).
 \label{eq:covariant-rate}
\end{equation}
The score bound supplies an explicit monotone method for optimizing
this structured family of achievable encoders.

\begin{proposition}[Alternating covariant updates]
\label{prop:covariant-alternating}
Suppose $a,b$ are strictly positive. Evaluate $\Omega_Q$ at
$p=p_{a,b}$ and form the noise score
\begin{equation}
 s_b(z)=\sum_va(v)\Omega_Q[p](v,vz),\qquad
 b^+(z)=\frac{b(z)e^{-s_b(z)}}{\sum_yb(y)e^{-s_b(y)}}.
 \label{eq:covariant-noise-update}
\end{equation}
Then $R(a,b^+)\geq R(a,b)+D(b\Vert b^+)$.
Next recompute the joint score at $p^+=p_{a,b^+}$ and set
\begin{equation}
 s_a(v)=\sum_zb^+(z)\Omega_Q[p^+](v,vz),\qquad
 a^+(v)=\frac{a(v)e^{-s_a(v)}}{\sum_w a(w)e^{-s_a(w)}}.
 \label{eq:covariant-outer-update}
\end{equation}
Then $R(a^+,b^+)\geq R(a,b^+)+D(a\Vert a^+)$.
Thus each coordinate step is monotone, and the alternating rate
sequence converges.
\end{proposition}

\begin{proof}
The change of variables $(v,z)\leftrightarrow(v,vz)$ is a bijection.
Consequently, with the other factor fixed,
\begin{align*}
 D(p_{a,\widetilde b}\Vert p_{a,b})&=D(\widetilde b\Vert b),&
 \langle p_{a,\widetilde b},\Omega_Q[p]\rangle
                                      &=\langle\widetilde b,s_b\rangle.
\end{align*}
The score bound~\eqref{eq:BN-Q} therefore yields
$F_Q(p_{a,\widetilde b})\leq
\langle\widetilde b,s_b\rangle+D(\widetilde b\Vert b)$,
with equality at $\widetilde b=b$. Gibbs minimization over
$\widetilde b$ gives~\eqref{eq:covariant-noise-update} and
\[
 F_Q(p_{a,b^+})\leq F_Q(p_{a,b})-D(b\Vert b^+).
\]
For the other coordinate use
$D(p_{\widetilde a,b^+}\Vert p_{a,b^+})=D(\widetilde a\Vert a)$
and the analogous averaged score. Negating these inequalities gives
the two rate bounds. Rates are bounded in finite dimension, so
monotonicity implies value convergence.
\end{proof}

Covariance also reduces the matrix computations in these steps. Set
$\rho_c=\sum_xc(x)\rho_x^B$, $\sigma_c=\sum_xc(x)\sigma_x^E$, and
\[
 d_c(x):=\Tr\sigma_x^E\log\sigma_c
                   -\Tr\rho_x^B\log\rho_c.
\]
Unitary conjugation of the conditional states gives the exact identity
\begin{equation}
 \Omega_Q[p_{a,b}](v,vz)=d_b(z)-d_{a*b}(vz).
 \label{eq:covariant-reduced-score}
\end{equation}
Thus $s_b(z)=d_b(z)-\sum_va(v)d_{a*b}(vz)$. For the outer
step, the analogous score is
$s_a(v)=\Delta(b^+)-\sum_zb^+(z)d_{a*b^+}(vz)$, where the
constant $\Delta(b^+)$ cancels upon normalization. These formulas
require matrix logarithms only for the averages associated with
the current noise law and its convolution, rather than separate
logarithms for every translated conditional state.

One may instead fix $a$ to be uniform and update only $b$. Since
$a*b$ is then uniform, the optimization target becomes
$\min_b\Delta(b)$, as in the uniform-outer-randomization bound of
\cite{SalekWiretap}. That paper proves equality with
$P^{(1)}(W)$ when the uniform input globally maximizes $\Delta$.
For a fixed blocklength
$n$, replacing $G$ by $G^n$ in Proposition~\ref{prop:covariant-alternating}
allows arbitrary correlations within both factors; the rate and the
improvement bounds are then divided by $n$. The contribution here is
the optimization rule and its descent certificate. The group-orbit
reduction, convolution identity, and operational coding and leakage
results are those of~\cite{SalekWiretap}.

\subsection{Numerical checks with commuting and noncommuting outputs}
\label{subsec:num-cq-erasure}

\begin{figure}[tb]
 \centering
 \includegraphics[width=.95\linewidth]{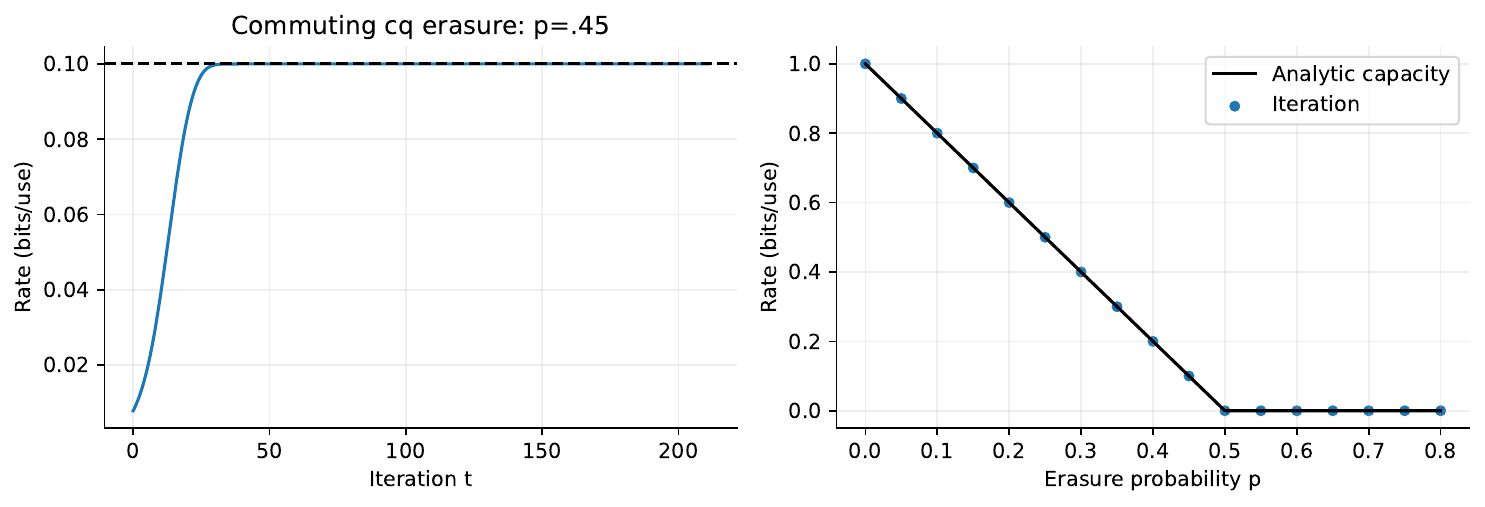}
 \caption{Commuting cq erasure check. Left: private information versus
 iteration at $p=0.45$. Right: numerical values and the exact capacity
 $\max\{0,1-2p\}$ across the erasure sweep. Rates are in bits.}
 \label{fig:cq-erasure-composite}
\end{figure}

For a commuting check, consider the binary-input erasure marginals
\[
 \rho_x^B=(1-p)|x\rangle\!\langle x|+p|e\rangle\!\langle e|,
 \qquad
 \sigma_x^E=p|x\rangle\!\langle x|+(1-p)|e\rangle\!\langle e|,
 \quad x=0,1.
\]
The flag $|e\rangle$ is orthogonal to $|0\rangle,|1\rangle$.
For every encoder,
$I(U;B)-I(U;E)=(1-2p)I(U;X)$, so the secrecy capacity is
\begin{equation}
 C_s(p)=\max\{0,1-2p\}\quad\text{bits/use}.
 \label{eq:Cs-cq-erasure}
\end{equation}
For $p<1/2$, uniform $U=X$ is optimal. For $p>1/2$, an independent
$U$ and $X$ attains zero; within the restriction $U=X$, a
deterministic input attains zero. Figure~\ref{fig:cq-erasure-composite}
checks the commuting specialization, evaluated in the common
eigenbasis, against this classical benchmark, including
$C_s(0.45)=0.10$ bits/use.

\subsection{A noncommuting channel requiring prefixing}
\label{subsec:num-cq-tikku}
We use the quantum-Bob family of Tikku, Berta, and
Renes~\cite[Section~V]{TikkuBertaRenes}:
\begin{equation}
 W_{BZ|X}[r]=\mathrm{BPC}(r)/\mathrm{BEC}((1-r)^2).
 \label{eq:tikku-family}
\end{equation}
Bob receives the pure states
$|\psi_0\rangle=|0\rangle$ and
$|\psi_1\rangle=r|0\rangle+\sqrt{1-r^2}|1\rangle$; Eve observes
the input through an erasure channel with erasure probability
$\alpha=(1-r)^2$. For $0<r<1$, Bob's states do not commute.
With $s=P_X(1)$, the no-prefixing objective in bits is
\begin{equation}
 f_r(s)=h_2\!\left(\frac{1+\sqrt{1-4s(1-s)(1-r^2)}}{2}\right)
            -\bigl[1-(1-r)^2\bigr]h_2(s).
 \label{eq:tikku-scalar}
\end{equation}
The binary convex-envelope formula gives
$P^{(1)}(W[r])=\max_s[f_r(s)-\operatorname{co}f_r(s)]$,
where $\operatorname{co}f_r$ is the lower convex envelope
\cite{TikkuBertaRenes,SalekWiretap}.
We compare the matrix update with a discretized convex-envelope
benchmark and with $\max_s f_r(s)$; the latter permits no prefixing.
The discretized benchmark gives a feasible lower approximation to $P^{(1)}$.

For this noncommuting example, we evaluate the quantum score by
eigendecomposition, flooring eigenvalues at $10^{-14}$ only when
taking logarithms, and evaluate the rate independently from
entropies. We use the stopping tests of
Section~\ref{subsec:numerical-protocol} with $\Omega_Q$ in place of
$\Omega$. The joint iteration uses 26 strictly positive initial laws at each
parameter: 12 random laws, nine perturbed diagonal laws, and five
symmetric prefixing laws. All retained best runs met the stopping
tests. At $r=0.542$, 11 of the 26 runs reached the cap of
$60\,000$ iterations; their values were below that of the retained
run. The envelope meshes have spacings $2\times10^{-4}$ and
$5\times10^{-5}$. Although their values differ by at most
$2.6\times10^{-9}$ bits across the sweep, the joint iteration exceeds
the finer-mesh value by up to $4.2\times10^{-7}$ bits near $r=0.51$.
A feasible mesh benchmark can miss small prefixing gains near the
boundary.

\begin{figure}[tb]
 \centering
 \includegraphics[width=\linewidth]{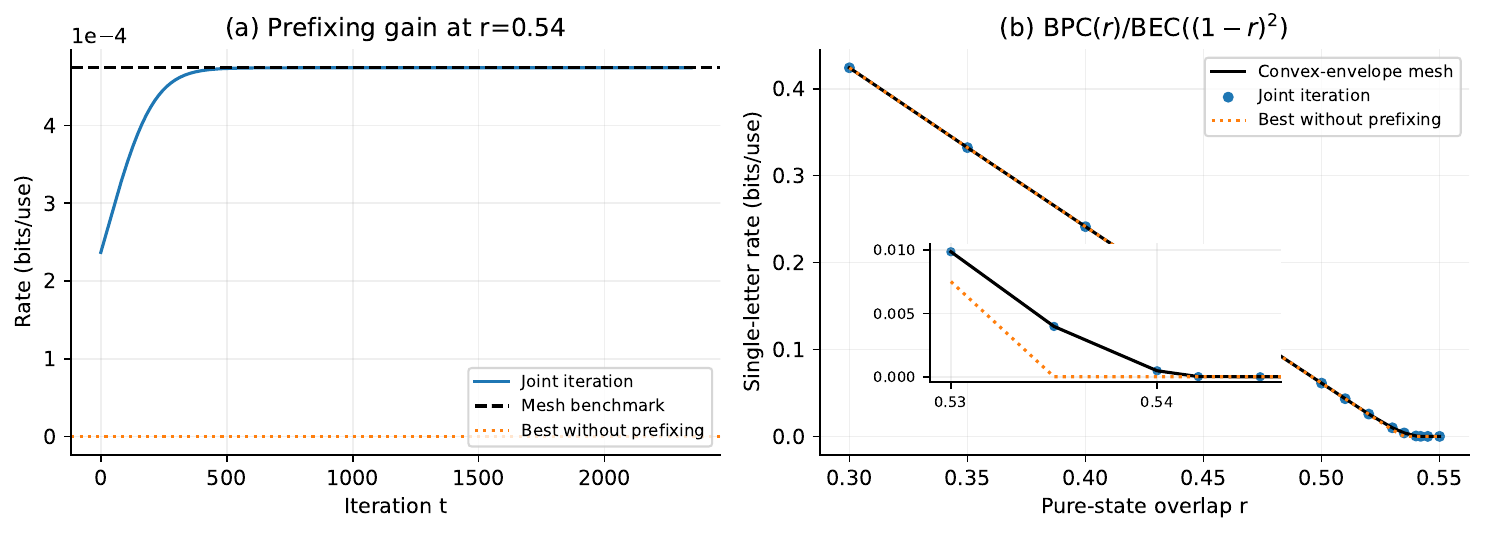}
 \caption{The noncommuting channel
 $\mathrm{BPC}(r)/\mathrm{BEC}((1-r)^2)$.
 Left: iteration at $r=0.54$ compared with the independent
 convex-envelope benchmark. Right: single-letter private information
 and the optimized no-prefixing objective across $r$.
}
 \label{fig:tikku-cqc-composite}
\end{figure}

At $r=0.54$, the numerical private information is approximately
$0.0004744481$ bits/use, whereas the no-prefixing maximum is zero.
A feasible encoder attaining the former value to the stated precision
has uniform $U$ and conditional input biases approximately
$0.23137$ and $0.76863$. The small positive rate therefore tests the
optimization over genuine stochastic prefixing, as well as the
noncommuting score evaluation. Figure~\ref{fig:tikku-cqc-composite}
shows the iteration and a parameter sweep. These are single-letter
private-information values; the capacity of this family can exceed
them through multi-use preprocessing~\cite{TikkuBertaRenes}.

We also tested Proposition~\ref{prop:covariant-alternating} on this
channel at $r=0.54$, using the $G=\mathbb Z_2$ action and 20 random
strictly positive factor pairs (random seed $20261012$). Every run
reached $0.00047444807086$ bits/use to the displayed precision;
one resulting pair was
\[
 a\simeq(0.5000000108,\,0.4999999892),\qquad
 b\simeq(0.2313705294,\,0.7686294706).
\]
The runs required $12\,757$--$19\,548$ full alternating cycles.
The stopping tests were applied to the induced joint law after
each full cycle, including the full-simplex first-order residual.
The largest observed decrease in a coordinate rate was below
$1.1\times10^{-15}$ bits, at floating-point precision. Thus the
structured update recovers the rate obtained by the unrestricted
joint iteration in this experiment.

\FloatBarrier

\subsection{A two-use encoder from the covariant wiretap construction}
We also apply the alternating update directly to the hybrid channel
of~\cite[Theorem~VIII.1]{SalekWiretap}, with Bob's channel
$\mathrm{BSC}(p)$, Eve's channel $\mathrm{BPC}(1-2p)$, and
$p=621/5000$. For $G^2=\mathbb Z_2^2$, that theorem uses the
explicit distributions
\[
 a_0=(0,\tfrac12,\tfrac12,0),\qquad
 b_0=(0.364,0.217,0.217,0.202),
\]
in the order $00,01,10,11$. Its two-use rate, divided by two, is
approximately $1.05086956\times10^{-6}$ nats/use. To meet the
strict-positivity assumption, we initialize
$a=(1-10^{-8})a_0+10^{-8}(1/4,1/4,1/4,1/4)$ and $b=b_0$.
After a fixed budget of $10\,000$ full alternating cycles, the
feasible rate is $2.45066123\times10^{-6}$ nats/use, or
$3.53555680\times10^{-6}$ bits/use. Coordinate rates are monotone
up to $5\times10^{-16}$ nats/use of floating-point roundoff.
This experiment improves the particular explicit witness used in
that theorem; it does not establish convergence at this budget,
global optimality, or an improvement over that paper's optimized
numerical benchmark. The nonadditivity proof itself remains the
result of~\cite{SalekWiretap}.

\FloatBarrier
\section{Discussion}
\label{sec:discussion}

The wiretap-specific update~\eqref{eq:F-map} jointly updates the
auxiliary distribution and stochastic prefixing. The
bound~\eqref{BN} proves the quantitative descent inequality
\eqref{eq:F-drop} for arbitrary finite classical channel pairs
without degradedness or less-noisy assumptions. The corresponding
argument extends to finite-dimensional cq outputs. The classical
and noncommuting examples with zero no-prefixing rate make the
role of stochastic preprocessing operationally visible.

The convergence guarantees depend on the channel assumptions.
The general theorem guarantees value convergence and stationarity
on each accumulation point's support face. Strictly positive
classical transition matrices give convergence of the entire
sequence through the square-root parametrization and the
Kurdyka--\L{}ojasiewicz inequality. Less noisiness gives a stronger
global conclusion for the restricted $U=X$ problem.

Random restarts and symmetry-informed initializations are useful
practical complements to monotonicity. Their attained rates remain
achievable lower bounds unless a separate global upper bound is
available. The finite-resolution convex-envelope calculations provide
independent small-alphabet benchmarks.

For covariant channels, Proposition~\ref{prop:covariant-alternating}
provides an algorithmic counterpart to the structural and operational
results of Hayashi and Salek~\cite{SalekWiretap}. Their group
convolution parametrization reduces the encoder search to two
distributions; the present score construction supplies explicit
monotone updates for those distributions. At blocklength $n$, the
same rule allows correlated distributions on $G^n$. Whether such a
search reaches the best structured encoder, or whether that family
achieves the unrestricted block optimum, remains a separate
question. Nonadditivity requires a rate per use above the optimized
one-use private information.

Useful next steps include sharper stopping certificates, adaptive
steps based on channel-specific contraction coefficients, and
experiments at larger input alphabets and blocklengths. The
unconditional score bound and the precise convergence distinctions
established here provide a basis for these developments.

\section*{Acknowledgments}
The authors are grateful to Debbie Leung and Graeme Smith for helpful
discussions.
F.S. acknowledges support from the European Commission through a
Marie Sk\l{}odowska-Curie Global Fellowship, and the hospitality of
Freie Universit\"at Berlin, Perimeter Institute for Theoretical Physics,
and the Institute for Quantum Computing, University of Waterloo.
M.H. acknowledges support in part from the Guangdong Provincial
Quantum Science Strategic Initiative (Grant No.~GDZX2505003) and
the General R\&D Projects of the 1+1+1 CUHK--CUHK(SZ)--GDST
Joint Collaboration Fund (Grant No.~GRDP2025-022).
The main result was obtained well before the widespread use of AI
tools. More recently, the authors used AI tools to assist with manuscript
preparation and numerical implementation.

\end{document}